\documentstyle[11pt,conf_iap]{article}

\include{psfig}

\def\Msun{{$M_\odot $}}
\def\Lsun{{$L_\odot $}}
\def\Teff{{$T_{ef\!f} $}}
\def\th{{\thinspace}}
\def\ni{\noindent}

\begin{document}

\heading{
MODELLING THE PHASE-LOCKED PULSATIONS OF A P=14d  EROS STAR}

% \photo{ }

 \author{Z. KOLL\'ATH$^{1,2}$, J.R. BUCHLER$^{1}$, J.P. BEAULIEU$^{3,4}$,
M.J. GOUPIL$^{5}$}
       {$^{1}$ Physics department, University of  Florida, Gainesville, FL   
       32611, USA.\\
       $^{2}$ Konkoly Observatory, 1025 Budapest, Box 67, Hungary. \\
       $^{3}$ Institut d'Astrophysique de Paris, CNRS, 98bis Boulevard Arago, 
       F--75014 Paris, France. \\
       $^{4}$ Kapteyn Laboratorium, Postbus 800, 9700 AV Groningen, The
       Netherlands. \\
       $^{5}$ DASGAL, Observatoire de Paris, Meudon 92195, France. }

 \bigskip
 
\begin{abstract}{\baselineskip 0.4cm 

 The Fourier spectral analysis \cite{poster1} of this object suggests a
nonlinear pulsation of period $\approx$\th 14d with a 3:2 frequency lock
between two vibrational modes.  We discuss an extensive search that we have
performed in the $L$, $M$, \Teff\ space for models of period 14--16 days that
exhibit this particular resonance.  We find models satisfying the resonance
constraints both in the Cepheid and in the post AGB regimes.  The nonlinear
pulsations of several candidates are presented.  The nature of the star remains
a mystery as none of the models satisfies all the observational constraints.
 }
 \end{abstract}

%def\psfig{psfig{\null}}

\section { Linear models }
 
We have performed an extensive search of linear nonadiabatic radial models with
periods of $P$=14--16 days.  We have chosen this value because nonlinear and
nonadiabatic effects tend to reduce the period somewhat, but the results are
insensitive to the exact value of the period.

The hydrostatic models and their linear stability analyses are computed for
composition parameters $X$=0.35--0.90, each with $Z$=0.004, $Z$=0.01 and
$Z$=0.02.  The OPAL \cite{opal} and Alexander \& Ferguson \cite{af} low
temperature opacities are used.  Convection is ignored.
For a fixed value of the temperature and the luminosity we iterate the mass to
get the desired pulsation period of $\approx$ 15 days for the fundamental mode
or for the first overtone.  Repeating this procedure for different values of
$T$ and $L$ we obtain the period ratio ($P_1 / P_0$ or $P_2 / P_1$) as a
function of these two parameters.

We present the results for $X$=0.7 and $Z$=0.01 in Figure 1. Each curve has a
fixed luminosity ($L$=2,000\Lsun\ for the most left and $L$=12,000\Lsun\ for
the most right curve).  The mass decreases from the left to the right in each
sequence.  The highest mass is between 3 and 6 \Msun, on the right side, the
calculations on the blue side of the sequences are terminated for masses 0.3
and 0.6 \Msun\ when we encounter numerical problems due to the high 
luminosity to mass ratio.
The dots represent the unstable (fundamental mode) models. Two groups of
unstable stars are found. The first group around \Teff=5,000\th K is inside the
Cepheid instability strip while the second one gives the region for the
unstable post-AGB stars (for a general linear survey of PAGB stars see eg
\cite{gautschy}).

 \centerline{\psfig{figure=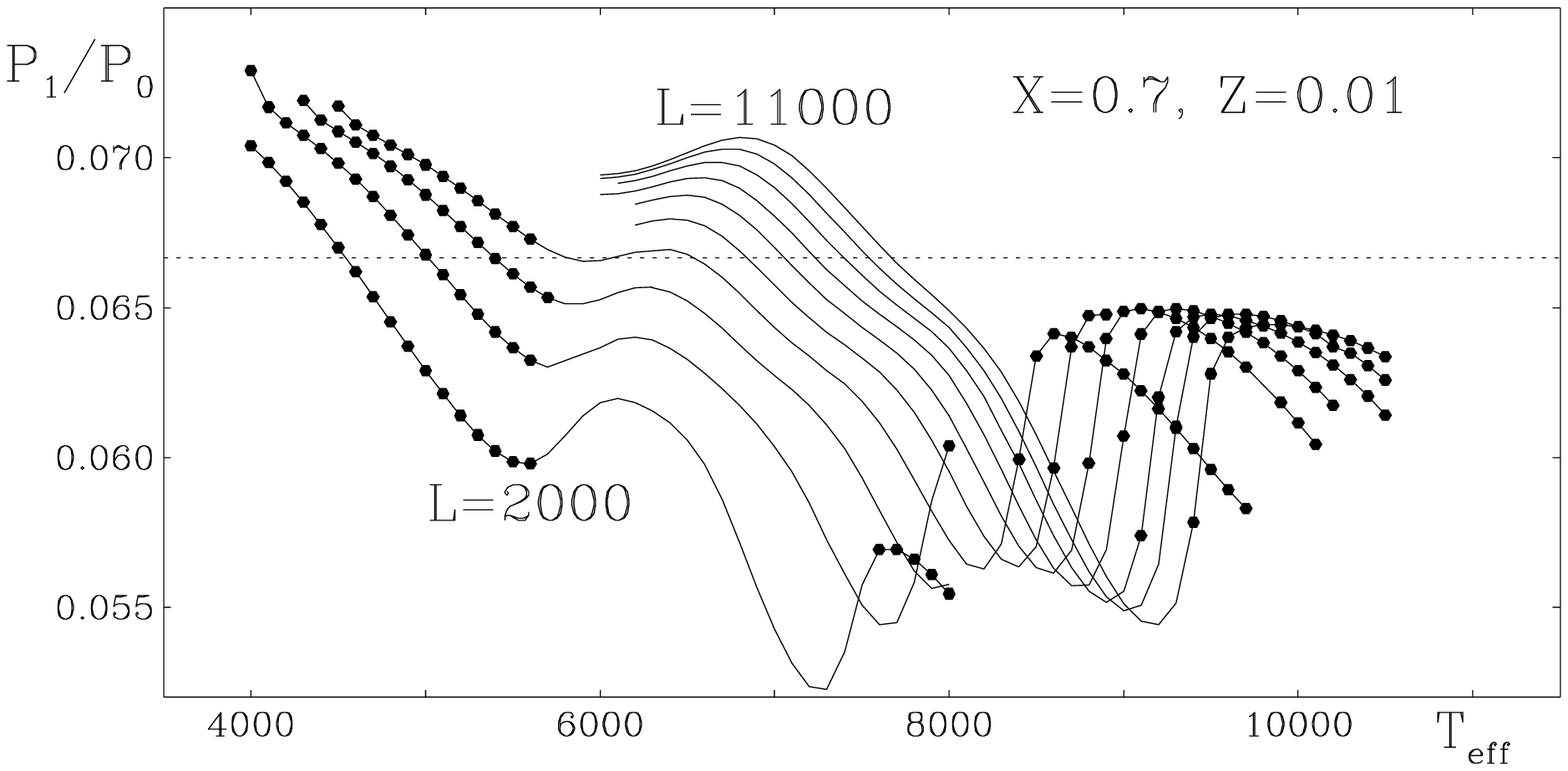,width=11cm}}

 \ni{\footnotesize FIG.~1.\ First overtone to fundamental mode period ratio as
a function of temperature, luminosity ranging from 2000 to 11000\Lsun; dots
denote linearly unstable models; on the left is the Cepheid group, on the right
the PAGB group.}
 
\vskip 15pt

Models with a 3:2 period ratio exist only among the Cepheids 
($L$=2,000--4,000\Lsun) for this composition of $X$=0.70, $Z$=0.01.  
However the observations
\cite{poster1} indicate a higher temperature and a higher luminosity for this
star than our search for these model parameters yield.

Interestingly, the period ratio increases close to the 3:2 resonance, right
were the instability occurs in the second, PAGB group.  The ratio $P_1/P_0$
reaches the highest value around \Teff=9,000\th K which, however, is still lower
than the spectroscopic estimate.

\vskip 30pt

 \centerline{\psfig{figure=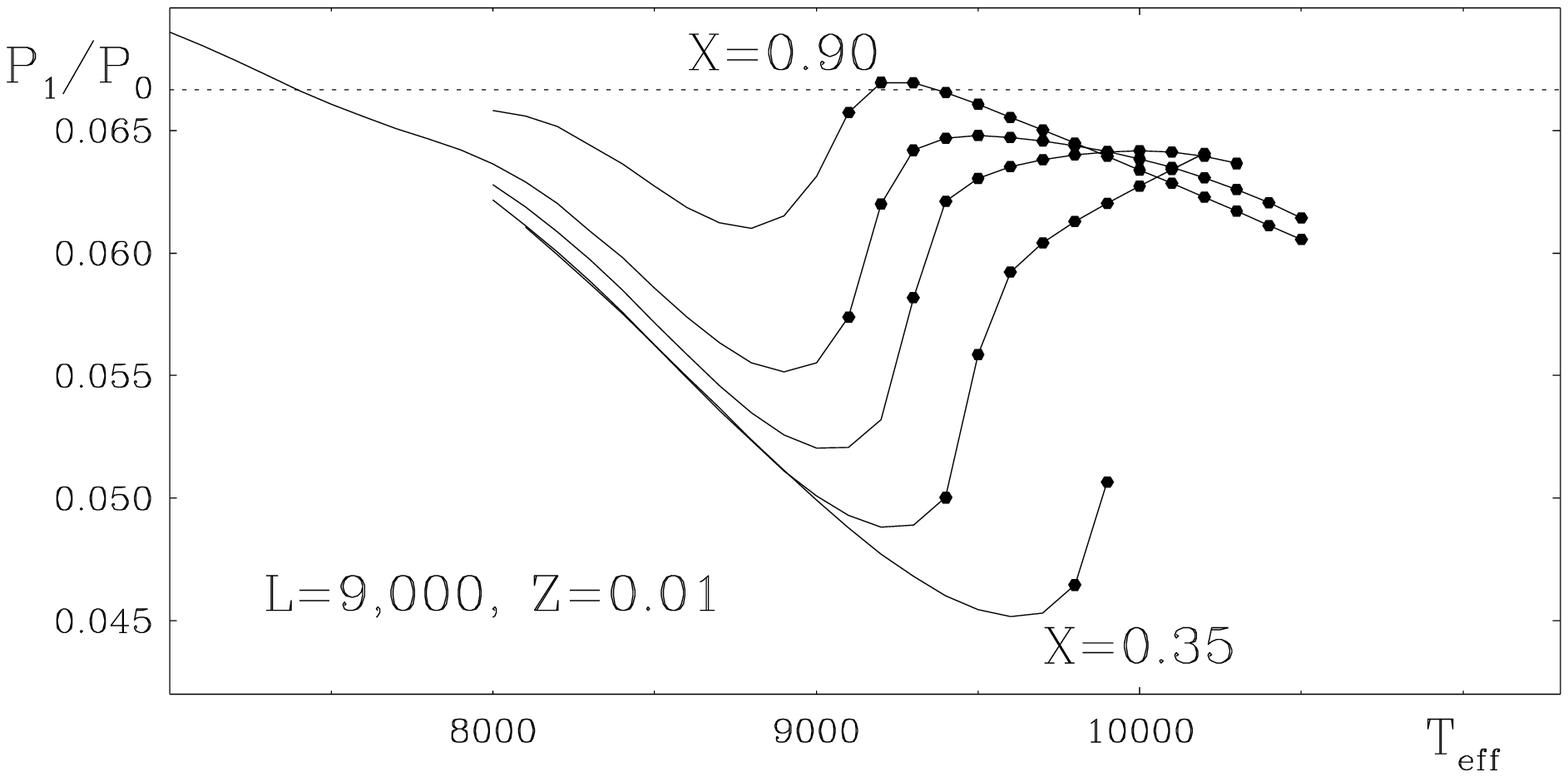,width=11cm}}

 \ni{\footnotesize FIG.~2.\ Period ratio ($P_1/P_0$) as a function of
temperature for different values of the hydrogen content ($X$).}

\vskip 40pt

 \centerline{\psfig{figure=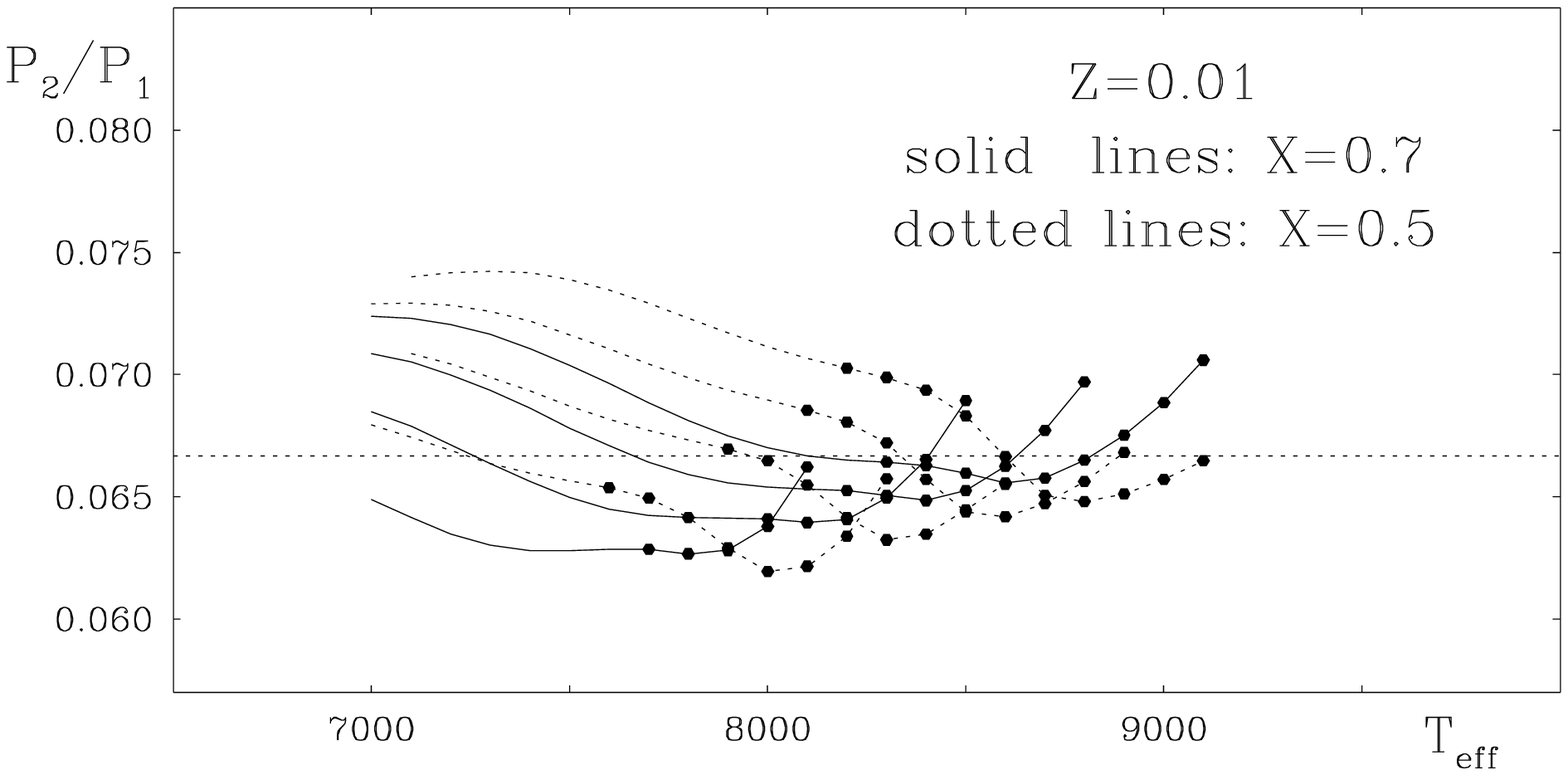,width=11cm}}

 \ni{\footnotesize FIG.~3.\ Period ratio ($P_2/P_1$) as a function of
temperature.  $L=$ 7000, 9000, 10000 and 13000 \Lsun\ (left to right).}

\vskip 15pt

The dependence of the period ratio on the composition is illustrated in Figure
2. Here we present only the high temperature models for $L$=9,000\Lsun. The
curves are given for the composition of $X$=0.9 (top), 0.7, 0.6, 0.5 and 0.35
(bottom) with $Z$=0.01.  There is a clear increase of the maximum value of the
period ratio as the hydrogen content is enhanced.  An increase to $X$=0.9 gives
the right period ratio, but this is an unacceptably large $X$, especially since
we are in the PAGB region.  For $X$=0.5 the period ratio is still increasing at
the blue end of the curve but both the mass and the linear growth rate become
small.

We have also performed tests with different values of metallicity.  Different
values of $Z$ only bring about a small shift of the curves.  We cannot push up
the period ratio to reach the resonance by varying the metallicity in the case
of normal $X$ values.

\vskip 10pt

We now turn to first overtone pulsators.  Here, there now also 
are solutions for the $P_2/P_1$=2/3 resonance for lower values of $X$.  
The period ratio as the function of the effective temperature is 
displayed on Figure 3.  The curves are given for $L$=7000, 9000, 11000 
and 13000\Lsun\ and for $X$=0.5 (dotted lines) and $X$=0.7 (solid lines). 
The resonance occurs in the range \Teff=8000--9000\th K.  The loci of the 
resonance are slightly shifted to the higher temperatures for the 
$X=$0.5 case compared to the higher $X$ models.

\section{Nonlinear models}

   From the Cepheid instability strip we have selected a resonant model with
the following parameters: $M$=3.8\Msun, $L$=3081\Lsun, \Teff=5156\th K, 
$X$=0.7 and $Z$=0.01\ with $P_0$=14.1\th d.
The numerical hydrodynamics shows that the resonance does indeed lead to a
phase-lock with alternating cycles ({\sl cf} \cite{MB} for an explanation of
this phenomenon).  The theoretical velocity and light curve are displayed on
Figure 4.  The light variation shows similar alternations as the EROS star, and
the Fourier amplitude spectrum is very similar to the observed one.  The
amplitude of the variation ($\approx$ 1.8 mag) is however much higher than the
observed value ($\approx$ 0.5 mag).  These differences are perhaps not
astonishing since both the temperature and luminosity of the model are much
lower than the observationally estimated values of the star.

\vskip 15pt

 \centerline{\psfig{figure=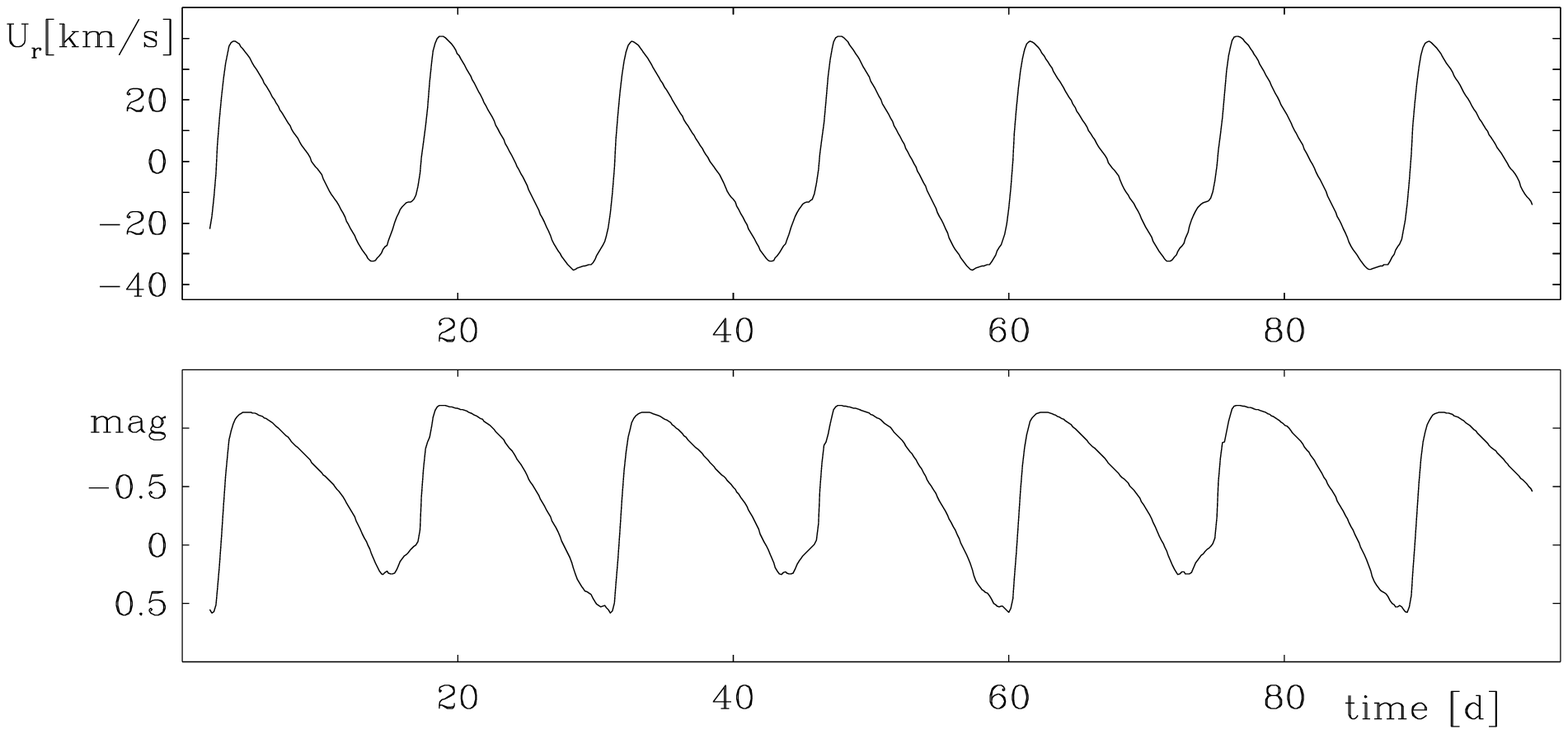,height=7cm,width=15cm}}

 \ni{\footnotesize FIG.~4.\ Variation of the Cepheid model. Velocity (top) 
 and light curve (bottom).}

\vskip 15pt

We have also calculated nonlinear models of the PAGB stars close to the linear
resonance.  In general we find that the computed amplitudes of these models are
very small (usually between 0.001--0.010 mag.) in agreement with calculations
of PAGB models \cite{zalewski}, but in disagreement with the observed value for
this star.  In contrast to the Cepheid model the growth rates are very large
and the models tend to undergo irregular (chaotic) pulsations.

One of the few models for which the linear frequencies satisfy closely the
resonance criterion has the following parameters: $M$=1.1406\Msun,
$L$=9566\Lsun, \Teff=9300\th K, $X$=0.9 and $Z$=0.01.  For this model both the
fundamental mode and the first overtone are linearly unstable (with the growth
rates of $\eta_0$=0.36 and $\eta_1$=0.37).  However, for none of the PAGB
models have we found any alternations similar to the observations.

\vskip 10pt

As far as overtone pulsators are concerned the linear results have indicated that
the $P_2$ to $P_1$ resonance provides good candidates for modelling this unique
star. However, the fundamental mode is also unstable, and all of our nonlinear
models (initialized with the first overtone velocity eigenvector) converge to
the fundamental mode pulsation after a short transient.

\section{Conclusion}

The nature of the $P$=14d phase locked star remains a mystery in spite of a
thorough search with radiative linear and nonlinear stellar models.  The linear
study shows that both Cepheid models and PAGB models have the right period
ratio between the fundamental mode and first overtone.  However, all of the
Cepheid models are underluminous and too cold compared to the observations.
Furthermore, numerical hydro computations show that the pulsation amplitudes
for these models are too high.  For PAGB models the resonance does not occur
with reasonable composition parameters for the fundamental mode, but can occur
for first overtone pulsators.  However, our hydrodynamics models are not able
to reproduce the observed alternations in the pulsations.  The theoretical
light curves of the high luminosity and high temperature models also have very
low amplitudes contrary to the observations.

It is important to get further observational confirmation and a better
estimation of the parameters of this unique variable star, because the
modelling can be an important test for our understanding of the physics inside
this kind of stars.

\acknowledgements{This work has been supported by the CNRS, the IAP and NSF.
 }

\vfill
\end{document}